\documentclass{article}

\usepackage{arxiv}

\usepackage{newtxtext,newtxmath}
\usepackage{arydshln}
\usepackage{float} 
\usepackage{comment}
\usepackage{graphicx}
\usepackage{xurl}
\usepackage{tikz}
\usetikzlibrary{positioning, arrows.meta, shapes.geometric}
\usepackage{amsmath}
\usepackage{mathtools}

\usepackage{endnotes}
\let\footnote=\endnote

\usepackage{natbib}
 \bibpunct[, ]{(}{)}{,}{a}{}{,}%

\usepackage{hyperref} 
\hypersetup{
	colorlinks,
	linkcolor={red!50!black},
	citecolor={blue!50!black},
	urlcolor={blue!80!black}
}

\begin{document}
%%%%%%%%%%%%%%%%

% Outcomment only when entries are known. Otherwise leave as is and
%   default values will be used.
%\setcounter{page}{1}
%\VOLUME{00}%
%\NO{0}%
%\MONTH{Xxxxx}% (month or a similar seasonal id)
%\YEAR{0000}% e.g., 2005
%\FIRSTPAGE{000}%
%\LASTPAGE{000}%
%\SHORTYEAR{00}% shortened year (two-digit)
%\ISSUE{0000} %
%\LONGFIRSTPAGE{0001} %
%\DOI{10.1287/xxxx.0000.0000}%

% Author's names for the running heads
% Sample depending on the number of authors;
% \RUNAUTHOR{Jones}
% \RUNAUTHOR{Jones and Wilson}
% \RUNAUTHOR{Jones, Miller, and Wilson}
% \RUNAUTHOR{Jones et al.} % for four or more authors
% Enter authors following the given pattern:
% \RUNAUTHOR{Bergman, et al.}

% Title or shortened title suitable for running heads. Sample:
% \RUNTITLE{Bundling Information Goods of Decreasing Value}
% Enter the (shortened) title:
% \RUNTITLE{U.S. Election Sensitivity}

% Full title. Sample:
% \TITLE{Bundling Information Goods of Decreasing Value}
% Enter the full title:
\title{The Sensitivity of the U.S. Presidential Election to Coordinated Voter Relocation}

\author{
	% You can write out first names or use initials - either way is acceptable, but be consistent
    C. Cardonha$^{1}$,
	D. Bergman$^{1}$,
    A. Cire$^{2}$,
	L. Lozano$^{3}$,
	T. Yunes$^{4}$ \and
	% Additional lines of authors should be inserted using the \and command (not \\)
	% Institution list, in a slightly smaller font
	\small$^{1}$ School of Business, University of Connecticut, Storrs, CT 06028, United States.\and
	\small$^{2}$Department of Management, University of Toronto Scarborough, Toronto, M1C 1A4, Canada.\and
	\small$^{3}$ Linder College of Business, University of Cincinnati, Cincinnati, OH 45221, United States.\and
	\small$^{4}$ Herbert Business School, University of Miami, Coral Gables, FL 33146, United States.\and
	% Identify at least one corresponding author, with contact email address
	% \small$^\ast$Corresponding author. Email: tallys@miami.edu \\
	% Joint contributions can be indicated like thisto this work.
}

\maketitle

% Block of authors and their affiliations starts here:
% NOTE: Authors with same affiliation, if the order of authors allows,
%   should be entered in ONE field, separated by a comma.
%   \EMAIL field can be repeated if more than one author
% \ARTICLEAUTHORS{%
% \AUTHOR{David Bergman}
% \AFF{Department of Operations and Information Management, University of Connecticut, Storrs, CT 06828, \EMAIL{david.bergman@uconn.edu}} %, \URL{}}
% \AUTHOR{Andre Cire}
% \AFF{Department of Management, University of Toronto Scarborough, Toronto, M1C 1A4, Canada, \EMAIL{m.arinella@adult.ufp.edu}}
% % Enter all authors
% } % end of the block

\begin{abstract} 
U.S. presidential elections are decided by the Electoral College, established in 1789, and designed to mitigate potential risks arising from the collusion of large groups of citizens. A statewide winner-take-all popular voting system for electors is implemented in all but two states, which has led to instances where narrow victories in key states were decisive in several recent elections. Small groups of voters can significantly impact the election, for example, through voter turnout. However, another dynamic can also influence this: a surprisingly small number of dedicated voters moving short distances across state lines. The extent to which the election's outcome is sensitive to small and well-coordinated movements of people has not been investigated in detail. Using a combination of forecasting, simulation, and optimization, we show that a candidate's probability of winning can be increased by 1\% through the strategic relocation of approximately 10,000 people no farther than 100 miles from their current county of residence, less than 0.006\% of the eligible voting population.  Moreover, an 8\% probability increase can be realized by a mere 50,000 voters relocating across state lines, or 0.03\% of the voting population. Given the remarkably small number of people involved and the fact that establishing electoral residence in many states takes about a month, this coordinated relocation of voters is not nearly as challenging as previously thought. As it stands, U.S. presidential elections may be vulnerable to the exploitation of the aforementioned loophole. Therefore, we anticipate our findings will have direct consequences on policymaking and campaign strategy, as well as motivate new operations research methods within the political sciences.
% The U.S. presidential election is decided by an Electoral College system, created in 1789 to prevent large groups from imposing proposals against the country's best interests. A winner-takes-all allocation of electoral votes in all but two states has led to narrow margins of victory in recent elections. Using forecasts of votes adjusted with data from recent polls and network-flow models, we show that small voter groups can significantly impact election outcomes. For example, relocating just 10,000 people within 100 miles could increase a candidate's win probability by 1\%. An almost 8\% increase can be achieved by a mere 50,000 voters relocating across state lines. These findings suggest that the election system may be vulnerable to strategic exploitation, with implications for policymaking, legislation, and campaign strategies.
\end{abstract}
% Enter your abstract

% Sample
%\KEYWORDS{deterministic inventory theory; infinite linear programming duality;
%  existence of optimal policies; semi-Markov decision process; cyclic schedule}

% Fill in data. If unknown, outcomment the field
% \KEYWORDS{Elections $<$ Government, Networks/graphs, Voting/committees $<$ Games/group decisions} \HISTORY{This paper was
% first submitted on XXX and has been with the authors for
% XXX years for XXX revisions.}

% \maketitle
%%%%%%%%%%%%%%%%%%%%%%%%%%%%%%%%%%%%%%%%%%%%%%%%%%%%%%%%%%%%%%%%%%%%%%

% Samples of sectioning (and labeling) in OPRE
% NOTE: (1) \section and \subsection do NOT end with a period
%       (2) \subsubsection and lower need end punctuation
%       (3) capitalization is as shown (title style).
%
%\section{Introduction.}\label{intro} %%1.
%\subsection{Duality and the Classical EOQ Problem.}\label{class-EOQ} %% 1.1.
%\subsection{Outline.}\label{outline1} %% 1.2.
%\subsubsection{Cyclic Schedules for the General Deterministic SMDP.}
%  \label{cyclic-schedules} %% 1.2.1
%\section{Problem Description.}\label{problemdescription} %% 2.

% Text of your paper here

\section{Introduction}

Could a small, coordinated interstate movement of voters influence the outcome of a U.S. presidential election? While this idea is often dismissed due to perceived legal, financial, and logistical challenges, we demonstrate that under current political conditions, even a modest, strategically coordinated movement of voters could significantly impact the 2024 election. To put this into perspective, relocating just half a football stadium's worth of voters a mere 100 miles could increase one candidate's chance of winning by nearly 8\%---a potential game-changer that could determine who sits in the Oval Office.

Our analysis is rooted in the intricate dynamics of the Electoral College,  where in all but two states (Maine and Nebraska) a winner-takes-all allocation of the electoral votes from that state is almost always implemented.
% , and all electoral votes are awarded to the candidate securing the majority of the popular vote in that state. 
%In the Electoral College system, established by the Founding Fathers in 1787, all but two states (Maine and Nebraska) award electors in a \textbf{winner-takes-all} basis to the candidate who secures the majority of the popular vote in that state. 
The power of the states in determining the outcome of a presidential election is heterogeneous~\citep{rabinowitz1986}, and the current electoral system amplifies the importance of \textit{swing} states. Over the past 25 years, elections have been determined by remarkably narrow differences in the popular vote within these states. One notable example is the 2000 presidential election, won by George W.\ Bush with 271 electoral votes (versus 266 for Al Gore). 
% In that election, the Republican candidate won the popular vote -- and, with that, 25 electoral votes -- in Florida by 537 votes,
% a mere  0.005\% of the eligible voters in that state~\citep{results2000}.
In that election, the Republican candidate won Florida’s 25 electoral votes by a margin of 537, representing just 0.005\% of eligible voters in the state~\citep{results2000}.

%\footnote{\url{https://www.fec.gov/resources/cms-content/documents/FederalElections2000_PresidentialGeneralElectionResultsbyState.pdf}}. %For additional background on the Electoral College and the power of the states in U.S. presidential elections, see \cite{rabinowitz1986, jameslawson1999}.

%\footnote{\url{{https://press.princeton.edu/books/ebook/9781400824281/breaking-the-deadlock?srsltid=AfmBOorZGJ4Px5Rw0IYju0PlzuQszBrEUTZUlWBPuh4MnxgAFzqim2ti}}. 

Beyond these thin margins of victory, another key factor to consider is the geographic proximity between swing states and states with more predictable outcomes. 
% (fig.~\ref{fig:movement-probability}).
%Take, 
For example, the neighboring swing states Nevada and Arizona
%, which 
share a border with California, a Democratic stronghold. Similarly, the swing state of Georgia is bordered almost entirely by Republican strongholds Tennessee, North Carolina, South Carolina, Florida, and Alabama. Referencing the 2000 election again and assuming one had access to perfect information, relocating 538 Democratic voters from Alabama or Georgia to Florida would have changed the outcome of that presidential election.

%Movement of people and its impact on voting outcome is nothing new. 
%The annual number of inter-state movements in the U.S. can exceed over 8 million people\cite{8MMovers}.  
%As an example, an article in the Washington Post\cite{ImpInterstate} discusses the implication of inter-state movements, where a desire to purposely block voters moving into states from the non-dominating party in the state is discussed. 
%Similarly, a recent article in The Washington Post\cite{2024movementVoters} discusses the party lines of voters that moved since the last presidential election. 
%This not only applies to the U.S., but countries globally, where researchers have studied party policies and how movement impacts voting\cite{CHOU_DANCYGIER_2021}. 
%Moving can and does impact voting\textemdash this paper studies how it can be coordinated in the case of the 2024 U.S. Presidential Election, with a significant and measurable impact even when involving a relatively modest number of people. Interestingly, our results suggest that 
%provides additional support for this concern, as it 

% this is just repeating the next sentence...
% Our study shows that a relocation strategy within this is strikingly simpler than one might think. 

% original:
% Through a combination of forecasting, simulation, and optimization, we show that a well-coordinates moves of small sizes and short distances could have a significant and measurable impact on the 2024 presidential elections.

Building on these two factors and through a combination of forecasting, simulation, and optimization, we demonstrate how precisely coordinated, small-scale, short-distance moves could dramatically influence the outcome of the 2024 election and possibly future ones. Our analysis is grounded in data-driven methods based on historical data on existing polls, showing that these subtle yet strategic adjustments are not only feasible but also significant in altering the electoral landscape.

Our work contributes to the existing research on elections in the operations research and the political science literature.
%For an overview of the O.R.\ side, we refer the interested reader to~\url{https://pubsonline.informs.org/editorscut/elections}. 
Specifically, U.S. elections have been studied since the 1960's \citep{hess1965nonpartisan, garfinkel1970optimal}. Special attention has been dedicated to political districting and gerrymandering \citep{validi2022political, validi2022imposing, swamy2023multiobjective},
%; most papers focus on the US elections (\cite{validi2022political, validi2022imposing, swamy2023multiobjective}), although other geographies have also been studied (\cite{bozkaya2011designing}). 
%In the case of the US,
and to the location of polling places
\citep{haspel2005}.
Districting plays a more important role in local elections, though, as the state-level popular vote  is not affected by internal subdivisions (except for Maine and Nebraska). The interest in election forecasts has also grown significantly over the years \citep{hummel2014, kaplan2003new}, especially after numerous polls failed to predict the outcome of the 2016 election, won by Donald Trump \citep{wright2018surprising}. Additional applications of  optimization to political problems include the allocation of resources to maximize seats in parliament \citep{guney2018} and estimating the minimal fraction of the popular vote necessary to elect a president in the Electoral College \citep{belenky2008}.

\section{Analytical Framework and Models}

We identify potential voter relocation strategies in a three-step process that results in a stochastic formulation of the problem.
%~\ref{fig:conceptualmap}. 
First, we design a simulation model to produce scenarios of voter turnout by party and county using voting data from 2004 to 2020~\citep{ElectionDataRaw}. Next, we  calibrate the results of this model using a Bayesian-style update with predictions from Nate Silver's projection site \citep{silver2024}, which aggregates recent news and polls. Finally, we
apply an optimization model based on sample average approximation \citep{KleywegtEtal01} that processes scenarios from the simulation model to identify voter movements from county to county. If movements are done in unison and by voters already planning to vote the same way, our solution maximizes the estimated probability that either the Republican or Democratic candidate wins the election. For practicality, we ensure that relocation distances are limited to counties located at most 100 miles apart and 
%(b) the number of voters moving out of a county must be less than or equal to the minimum number of voters for a given party and county; 
that the destination county of a move must be located in one the following swing states: Arizona, Georgia, Michigan, Nevada, New Hampshire, North Carolina, Pennsylvania, and Wisconsin. These  were chosen because there was no candidate having more than a 70\% chance of winning these states when this study was conducted.
% Technical details of the method are provided in the supplemental material (SM).
%(fig.~\ref{fig:conceptualmap}).
The remainder of this section formalizes the problem and presents the three models, detailing the inputs and methodologies employed. %Finally, we discuss the validation measures we employed to ensure the simulation model achieved stable and reasonable scenarios of voter turnout.

\subsection{Problem Description}

\newcommand{\party}{\ensuremath{p}}
\newcommand{\units}{\ensuremath{\mathcal{U}}}
\newcommand{\unit}{\ensuremath{u}}
\newcommand{\nunits}{\ensuremath{n}}
\newcommand{\evotes}{\ensuremath{v}}
\newcommand{\pvotes}{\ensuremath{\tilde{N}}}
\newcommand{\states}{\ensuremath{\mathcal{S}}}

We investigate the problem from the perspective of the two main parties. We refer to them as $\party_1$ and~$\party_2$ in our model, and assume without loss of generality that we are solving the problem for party~$\party_1$.
% other parties and independent candidates are considered in aggregate.  
Let $\units$
%\coloneq \{\unit_1, \ldots, \unit_{\nunits}\}$ 
be the set of electoral units, which is a partition $\mathcal{U} := \mathcal{S} \cup \mathcal{D}$ of state units $\mathcal{S}
$ and district units $\mathcal{D}$. Each U.S. state has a one-to-one mapping to state units in $\mathcal{S}$. Further, Maine and Nebraska are associated with two and three district units, respectively, in $\mathcal{D}$.
Each  unit $\unit$ is assigned~$\evotes_\unit \in \mathbb{Z}_+$ electoral votes in the presidential election, with~$\sum_{\unit \in \units}\evotes_\unit = 538$. A candidate who wins the popular vote at a unit $u$ receives all $\evotes_u$ votes.

\newcommand{\counties}{\ensuremath{\mathcal{C}}}
\newcommand{\county}{\ensuremath{c}}

Each unit $\unit \in \units$ is composed of counties $\counties_u$, where $\counties \coloneqq \bigcup_{\unit \in \units} \counties_u$ is the set of U.S. counties. Each county belongs to exactly one state unit, and all votes in a county $\county \in \counties_s$ count towards the state $s$. A county $c$ in the states of Maine and Nebraska belongs to at least one district unit $C_d$, $d \in \mathcal{D}$. Let $f_{\county,d} \in [0,1]$ be the fraction of county $\county$'s population within district unit $d \in \mathcal{D}$. We assume that each vote in a county $c \in \counties_{d}$ counts for $f_{\county,d}$ votes towards the district
$d \in \mathcal{D}$.

% %= \{\county_1, \ldots, \county_l\}$ 
% be the set of U.S. counties and $\mathcal{C}_u \subseteq \mathcal{C}$ be the set of counties in unit $u$.  Each county belongs entirely within one state unit; we let $\mathcal{C}_s \subseteq \mathcal{C}$ be the set of counties in state $s$.  In Maine and Nebraska, each county belongs to at least one district, but is sometimes split between the district units. We let $f_{\county,d}$ be the fraction of county $\county$'s population within district unit $d \in \mathcal{D}$ and let $\mathcal{C}_d \subseteq \mathcal{C}$ be the set of counties in district unit $d$.

% \newcommand{\move}{\ensuremath{\mathbf{X}}}

\newcommand{\total}{\ensuremath{\tilde{T}}}
\newcommand{\win}{\ensuremath{\tilde{W}}}

We wish to identify a movement matrix $\mathbf{X} \in \mathbb{Z}_{\geq 0}^{|\counties| \times |\counties| }$ to most effectively increase the probability of~$\party_1$ winning, where $x_{i,j}$ is the number of people relocating from county $c_i$ to county $c_j$. Each movement only considers identifiable, highly engaged electors of~$\party_1$, i.e., they will vote in the elections and choose~$\party_1$. Specifically, let $\pvotes_{\party_1,\unit}$ and $\pvotes_{\party_2,\unit}$ be random variables representing the number of votes parties~$\party_1$ and $\party_2$ receive in unit $\unit$, respectively. We consider the following stochastic problem:
\begin{equation}
\begin{aligned}
\max_{\mathbf{X}} \quad & \mathbb{P} \left( \sum_{\unit \in \units} v_\unit \win_{\unit} \geq 270 \right) 
\\
\text{s.t.} \quad 
    & \win_{\unit} = \mathbb{I} \left(
\total_{\unit} > \tilde{N}_{p_2,\unit} \right) 
    && \forall \unit \in \units \\
\quad 
    & \total_{s} =
\pvotes_{\party_1,s} + \sum_{c_i \in \mathcal{C}_s} \sum_{c_j \in \mathcal{C} \setminus \mathcal{C}_s} \left( 
x_{j,i} -  x_{i,j}
\right)
    && \forall s \in \mathcal{S} \\
     & \total_{d} =
\pvotes_{\party_1,d} + \sum_{c_i \in \mathcal{C}_d} f_{c_i,d} \sum_{c_j \in \mathcal{C} \setminus \mathcal{C}_d} \left( 
x_{j,i} -  x_{i,j}
\right)
    && \forall d \in \mathcal{D}
\\ 
& \mathbf{X} \in \Omega.
\end{aligned}
\tag{\textbf{P}}
\label{eq:optimization_problem}
\end{equation}
In formulation~\ref{eq:optimization_problem}, $\total_{\unit}$ is a random variable that denotes the number of votes received by~$\party_1$ in unit~$\unit$ after the relocations in and out of all counties in~$\counties_\unit$, as defined in the second and third constraints. Party $\party_1$ obtains all $\evotes_{\unit}$ electoral votes if $\total_{\unit} > \tilde{N}_{\party_2,\unit}$, which here is modelled as a Bernoulli random variable $\win_{\unit}$ in the first constraint. The objective maximizes the probability that $p_1$ wins the election by receiving at least 270 electoral votes. Finally, the set~$\Omega$ contains the feasible solutions to~\ref{eq:optimization_problem}; we discuss the constraints derived from practical considerations in~\S\ref{sec: network model}.

\subsection{Voter Turnout Simulation Model}\label{sec:simulation model}

We simulate the number of votes received by each candidate in each of the 3,150 counties using growth models derived from county-level voting returns for the last five U.S. presidential elections (2004-2020) with data from the MIT Election Data Science Lab (\url{https://electionlab.mit.edu/}). Our statistical model is designed to capture a well-known spatial correlation across counties under partisan stability across the years; see, e.g., \cite{kim2003spatial, fiorino2022detecting,  2024movementVoters}.

%The simulator was designed to provide a scalable and reproducible model for voter turnout per candidate in each of the 3,150 counties in the dataset. 

Specifically, the number of voters per county per candidate is modeled as a collection of correlated lognormal distributions 
that defines the baseline for simulating voter turnout.  For each party $\party$, county $\county$, and election year $t$ (indexed by $1, 2, \ldots, 5$, where $1$ represents 2004, 2 represents 2008, and so forth), let $V_{\party,\county}$ be the logarithm of the number of people who voted for party $\party$ in county $\county$. We define the number of voters for party $\party$ in county $\county$ as
$
\tilde{V}_{\party,\county} \sim \mathcal{LN}\left(\mu_{\party,\county}, \sigma_{\party,\county}\right),$ i.e.,
  a lognormal distribution with mean $\mu_{\party,\county}$ and standard deviation $\sigma_{\party,\county}$.  

The lognormal distribution was chosen for four primary reasons: (a) it is positive; (b) captures counties with small voting populations; (c) correlates variables based solely on historical trends; and (d) accounts for heteroskedasticity observed in our data. We employed a Breusch-Pagan test for heteroskedasticity using 2020 county population data, which yields a Lagrange multiplier of 1,289.73, p-value $\le$ 0.0001, when fit to votes for the Democratic candidate. That is, the variance of the residuals is not constant when using the 2020 county population, thereby justifying the use of a distribution that can accommodate such variability. All voters for candidates other than the primary Republican or Democratic candidate are grouped into a single ``other'' category. Thus, our model consists of 9,450 lognormal random variables, where 6,300 variables are associated with the two main  candidates. Finally, we note that the discrete outcomes (number of votes) are typically sufficiently large so that a continuous distribution is still applicable.

 The simulation is parameterized by a vector~$\boldsymbol{\hat{\mu}}$ describing estimates for the expectations of each random variable and a covariance matrix~$\hat{\Sigma}$. We model voter turnout  using estimates of the average growth rate $\hat{g}_{\party,\county}$ for each party $\party$ and county $\county$ based on historical voter data, as follows:
\begin{eqnarray*}
\hat{g}_{\party,\county} \coloneqq \frac{1}{4} \sum_{t=1}^{4} \frac{\left(V_{\party,\county,t+1}\right) - \left(V_{\party,\county,t}\right)}{V_{\party,\county,t}}.    
\end{eqnarray*}

Based on the growth rates above, the expected turnout for 2024 is estimated as
\begin{eqnarray*}
\hat{\mu}_{\party,\county} \coloneqq \mathbb{E}\left[\log\left(\tilde{V}_{\party,\county}\right)\right] = V_{\party,\county,5} \cdot \hat{g}_{\party,\county}.
\end{eqnarray*}
We derive~$\hat{\Sigma}$ by computing the sample covariance of the log-transformed voter turnout from the past five elections. If the resulting matrix is not positive semi-definite, we replace it with the nearest symmetric positive semidefinite matrix within the Frobenius norm (\citealt{HIGHAM1988103}). 

Thus, given~$\boldsymbol{\hat{u}}$ and $\hat{\Sigma}$, we generate $z$ scenarios of voter turnout as follows:
\begin{enumerate}
    \item (Normal Data Simulation) 
    We sample~$z$ matrices $\mathbf{v}_1, \dots, \mathbf{v}_z$ from the distribution $\mathcal{N}(\mathbf{\hat{u}}, \hat{\Sigma})$. Each sample $v_{q,\party,\county}$ represents the log-transformed voter turnout of party $\party$ in county $\county$ sampled in scenario $q \in [z] \coloneqq \{1,\dots,z\}$. 
    \item (Lognormal Rescale) We exponentiate the adjusted normal data to recover the sampled voter turnout~$\xi'_{q,\party,\county}$ for each party, county, and scenario, i.e., $
   \xi'_{q,\party,\county} \coloneqq e^{v_{q,\party,\county}}
   $. 
\end{enumerate}

\smallskip 
The matrix $\mathbf{\Xi}' \in \mathbb{R}^{z \times 3 \times |\counties|}_{+}$ is the result of the process described above, which includes sampled data describing voting turnout in each scenario.

\smallskip
\noindent \textbf{Bayesian-style Update Model.} The simulation model is based on historical voter trends and does not account for up-to-date information available from recent polls and prediction services. To incorporate this data, we develop a Bayesian-style update to transform~$\mathbf{\Xi}'$ into a matrix~$\mathbf{\Xi}$ such that the proportion of simulations in which a party wins a state matches the given predictions.
%that reflects information of recent election data.
%We calibrate the results in~$\mathbf{\Xi}'$ to incorporate recent trends observed on recent pools and predictions through a Bayesian-style update.
%To calibrate to updated data based on polls and prediction sites, we transform $\mathbf{\Xi}'$ into $\mathbf{\Xi}$ through a Bayesian-style update. 
%To adjust our simulation based on recent election data, we utilize a Bayesian update to identify a multiplier for each electoral unit. 
%Such a multiplier switches voters between parties to best conform to recent state-wide predictions. 
Our analysis considers Nate Silver's state-by-state predictions as of August 18, 2024 \citep{silver2024} as our basis.
Other dates for data extraction were tested and produced similar results.

% \citeAppendix{silver2024}. 
%The average of these multipliers across the 51 states and special districts is 4.3\% (see table~\ref{tab:sim_results}). We present a more detailed description below.

\newcommand{\state}{\ensuremath{s}}

Formally, consider a state unit $s \in \mathcal{S}$ for which a party's exogenous winning probability $\theta_s$ (from Nate Silver's prediction) is greater than its winning probability in $\mathbf{\Xi}'$ by more than 0.01. We refer to such a party as the \textit{increasing party} $\party_{\uparrow}$, as its number of votes must increase for the winning probability to match $\theta_s$. The adversarial party, in turn, is referred to as the \textit{decreasing party} $\party_{\downarrow}$. Our goal is to switch~$\lambda_s$ percent of the voters from $\party_{\downarrow}$ to $\party_{\uparrow}$ in~$\mathbf{\Xi}'$. We use the following mixed-integer programming formulation to identify the minimum $\lambda_s$:
\begin{subequations}
\begin{align}
\lambda^*_s \coloneqq \min_{\lambda_s, \mathbf{w}} \quad &  \lambda_s \label{eq:1a}
\\
\textnormal{s.t.} \quad 
    & y^{\party_{\uparrow}}_{q} =
\sum_{\county \in \mathcal{C}_s} \xi'_{q,\party_{\uparrow},\county} + \lambda_s\sum_{\county \in \mathcal{C}_s} \xi'_{q,\party_{\downarrow},\county} && \forall q \in [z] \label{eq:1b} \\ 
\quad & y^{\party_{\downarrow}}_{q} =
(1-\lambda_s)\sum_{\county \in \mathcal{C}_s} \xi'_{q,\party_{\downarrow},\county} && \forall q \in [z] \label{eq:1c} \\
\quad & y^{p_{\uparrow}}_{q} +M(1-w_{q}) \geq y^{p_{\downarrow}}_{q} + w_{q} && \forall q \in [z]  \label{eq:1d} \\
\quad & \frac{1}{z}\sum_{q \in [z]} w_{q} \geq \theta_s  \label{eq:1e} \\
\quad & \lambda_s \in [0,1], w_q \in \{0,1\} && \forall q \in [z].  \nonumber 
\end{align}
\label{eq:bayesianUpdate}
\end{subequations}

In addition to variable~$\lambda_s$, which we wish to minimize in~\eqref{eq:1a}, the model uses auxiliary variables~$y_q^{\uparrow}$ and~$y_q^{\downarrow}$  to represent the updated number of votes for each party and county in the $q$-th scenario. Moreover, the model uses binary variables~$w_{q}$ to indicate whether or not the increasing party wins~$\state$ in scenario $q$. Constraints~\eqref{eq:1b} and~\eqref{eq:1c}
set the values of~$y_q^{\uparrow}$ and~$y_q^{\downarrow}$ based on~$\lambda_s$. Constraints~\eqref{eq:1d} set each binary variable~$w_{q} = 1$ if and only if~$p_\uparrow$ receives more votes than~$p_\downarrow$ in the $q$-th scenario; $M$ is a sufficiently large constant. Lastly, \eqref{eq:1e} asserts that our win probability estimate for~$\party_{\uparrow}$ matches the prediction~$\theta_\state$.

Given an optimal solution~$\lambda^*_\state$, we update the voter turnout scenarios in~$\mathbf{\Xi}'$ for state~$\state$ as follows:
\begin{align*}
   & \xi_{q,\party_{\uparrow},\county} = \xi'_{q,\party_{\uparrow},\county}+\lambda^*_s\xi'_{q,\party_{\downarrow},\county} &\forall \county  \in \mathcal{C}_s, \forall q \in [z] \\
   & \xi_{q,\party_{\downarrow},\county} = (1-\lambda^*_s)\xi'_{q,\party_{\downarrow},\county} &\forall \county \in \mathcal{C}_s, \forall q \in [z].
\end{align*}

Maine and Nebraska use a modified model to account for district units. Instead of defining a single multiplier for the whole state, we consider two multipliers per district, each denoting the switch of voters between the parties in one district to the other. For the constraints, we require that the updated estimated probability for the state and the districts be within a small tolerance of the exogenous winning probability. Finally, the objective is to minimize the sum of all the multipliers, again aiming to switch the least amount of people.

\begin{figure}[h]
    \centering
    \includegraphics[width=0.4\textwidth]{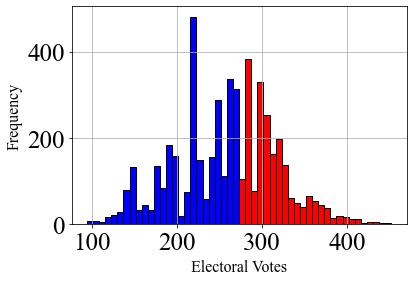} % Replace with your figure's path
    \caption{Electoral Vote Count for Republicans in 5,000 Simulations }
    \label{fig:electoralvotes-probability}
\end{figure}

\smallskip
\noindent \textbf{Validation of the Simulation Model.} 
We validate the accuracy of our simulation model by comparing our results with the estimates of the online prediction market platform \texttt{PredictIt}~\citep{predictit2024}.
%\footnote{\url{https://www.predictit.org/markets/detail/6867/Which-party-will-win-the-2024-US-presidential-election}}. 
At the time of data extraction, the presumptive Republican and Democratic candidates had a  44.66\% and 54.37\% chance of winning the general election, respectively, assuming a basic normalization on betting lines. In 5,000 scenarios from our simulation model, the Republican and Democratic candidates win 41.70\% and 58.02\% of the time, respectively, showing reasonable calibration with betting markets at the time of data extraction.  
The histogram of electoral votes for the 5,000 scenarios (Figure~\ref{fig:electoralvotes-probability}) shows a reasonable distribution of electoral votes. 
The electoral votes won by the Republican candidate have a first quartile (Q1) of 219, a median of 262, and a third quartile (Q3) of 301. The electoral votes won by the Democratic candidate have a Q1 of 237, a median of 276, and a Q3 of 319. The mean number of electoral votes is 257.95 and 280.05 for Republican and Democratic candidates, respectively.  The electoral-unit level pre-movement summary statistics of the scenarios are detailed in Table~\ref{tab:sim_results}.  Relevant data generated from our simulation model can be accessed at \url{https://www.dropbox.com/s/efwttk3xg21nh6z/ElectionsPaperData.zip?dl=1}.

\begin{table}[b!]
\tiny
\centering
\begin{tabular}{|c|c|c|c|c|c|c|c|}
\hline
\textbf{Electoral Unit} & \textbf{Electoral Votes} & \textbf{Total Votes (M)} & \textbf{Rep. Votes (M) } & \textbf{Dem. Votes (M) } & \textbf{Bayesian Factor ($\lambda^*_\state$)} & \textbf{Rep. Win Prob.} & \textbf{Dem. Win Prob.} \\
\hline
AL                      & 9                        & 2.442                    & 1.527 (1.354,1.736)     & 0.915 (0.794,1.058)     & 0.000                    & 100\%                   & 0\%                     \\ \hline
AK                      & 3                        & 0.349                    & 0.185 (0.120,0.302)     & 0.165 (0.122,0.229)     & 0.014                    & 81\%                    & 19\%                    \\ \hline
AZ                      & 11                       & 3.849                    & 1.920 (1.524,2.387)     & 1.929 (1.421,2.571)     & 0.028                    & 50\%                    & 50\%                    \\ \hline
AR                      & 6                        & 1.251                    & 0.822 (0.698,0.970)     & 0.429 (0.379,0.508)     & 0.000                    & 100\%                   & 0\%                     \\ \hline
CA                      & 54                       & 18.967                   & 6.245 (5.160,7.508)     & 12.723 (9.861,16.204)   & 0.000                    & 0\%                     & 100\%                   \\ \hline
CO                      & 10                       & 3.572                    & 1.703 (1.445,2.009)     & 1.868 (1.423,2.421)     & 0.119                    & 10\%                    & 90\%                    \\ \hline
CT                      & 7                        & 1.879                    & 0.868 (0.792,0.953)     & 1.011 (0.862,1.179)     & 0.125                    & 2\%                     & 98\%                    \\ \hline
DE                      & 3                        & 0.543                    & 0.251 (0.214,0.295)     & 0.291 (0.236,0.358)     & 0.122                    & 6\%                     & 94\%                    \\ \hline
DC                      & 3                        & 0.374                    & 0.019 (0.013,0.027)     & 0.355 (0.278,0.448)     & 0.000                    & 0\%                     & 100\%                   \\ \hline
FL                      & 30                       & 12.164                   & 6.215 (4.972,7.724)     & 5.949 (4.835,7.277)     & 0.007                    & 81\%                    & 19\%                    \\ \hline
GA                      & 16                       & 5.617                    & 2.827 (2.428,3.311)     & 2.790 (2.137,3.608)     & 0.057                    & 62\%                    & 38\%                    \\ \hline
HI                      & 4                        & 0.628                    & 0.263 (0.197,0.351)     & 0.366 (0.285,0.463)     & 0.130                    & 4\%                     & 96\%                    \\ \hline
ID                      & 4                        & 0.942                    & 0.606 (0.492,0.741)     & 0.336 (0.258,0.429)     & 0.000                    & 100\%                   & 0\%                     \\ \hline
IL                      & 19                       & 6.193                    & 2.796 (2.479,3.148)     & 3.398 (2.963,3.902)     & 0.081                    & 1\%                     & 99\%                    \\ \hline
IN                      & 11                       & 3.199                    & 1.856 (1.591,2.165)     & 1.343 (1.061,1.682)     & 0.031                    & 98\%                    & 2\%                     \\ \hline
IA                      & 6                        & 1.734                    & 0.951 (0.798,1.125)     & 0.783 (0.655,0.937)     & 0.000                    & 89\%                    & 11\%                    \\ \hline
KS                      & 6                        & 1.415                    & 0.767 (0.693,0.847)     & 0.648 (0.533,0.783)     & 0.023                    & 99\%                    & 1\%                     \\ \hline
KY                      & 8                        & 2.225                    & 1.411 (1.203,1.659)     & 0.814 (0.707,0.942)     & 0.000                    & 100\%                   & 0\%                     \\ \hline
LA                      & 8                        & 2.188                    & 1.134 (1.050,1.242)     & 1.054 (0.990,1.128)     & 0.135                    & 98\%                    & 2\%                     \\ \hline
ME                      & 2                        & 0.834                    & 0.386 (0.331,0.448)     & 0.448 (0.385,0.517)     & 0.000                    & 20\%                    & 80\%                    \\ \hline
ME Dist. 1                   & 1                        & 0.454                    & 0.177 (0.158,0.197)     & 0.277 (0.228,0.338)     & 0.034                    & 0\%                     & 100\%                   \\ \hline
ME Dist. 2                   & 1                        & 0.379                    & 0.209 (0.173,0.251)     & 0.170 (0.138,0.207)     & 0.000                    & 87\%                    & 13\%                    \\ \hline
MD                      & 10                       & 3.182                    & 0.972 (0.920,1.042)     & 2.210 (1.794,2.702)     & 0.000                    & 0\%                     & 100\%                   \\ \hline
MA                      & 11                       & 3.764                    & 1.199 (1.103,1.300)     & 2.565 (2.155,3.034)     & 0.000                    & 0\%                     & 100\%                   \\ \hline
MI                      & 15                       & 5.712                    & 2.796 (2.370,3.293)     & 2.916 (2.467,3.428)     & 0.011                    & 37\%                    & 63\%                    \\ \hline
MN                      & 10                       & 3.360                    & 1.616 (1.468,1.789)     & 1.744 (1.501,2.009)     & 0.044                    & 9\%                     & 91\%                    \\ \hline
MS                      & 6                        & 1.370                    & 0.714 (0.660,0.780)     & 0.656 (0.556,0.778)     & 0.093                    & 94\%                    & 6\%                     \\ \hline
MO                      & 10                       & 3.086                    & 1.807 (1.604,2.056)     & 1.279 (1.053,1.561)     & 0.000                    & 99\%                    & 1\%                     \\ \hline
MT                      & 4                        & 0.646                    & 0.356 (0.291,0.432)     & 0.289 (0.230,0.359)     & 0.035                    & 96\%                    & 4\%                     \\ \hline
NE                      & 2                        & 0.998                    & 0.577 (0.503,0.659)     & 0.421 (0.340,0.520)     & 0.000                    & 100\%                   & 0\%                     \\ \hline
NE Dist. 1                   & 1                        & 0.349                    & 0.195 (0.167,0.227)     & 0.154 (0.123,0.191)     & 0.000                    & 97\%                    & 3\%                     \\ \hline
NE Dist. 2                   & 1                        & 0.350                    & 0.154 (0.139,0.170)     & 0.196 (0.148,0.256)     & 0.035                    & 1\%                     & 99\%                    \\ \hline
NE Dist. 3                   & 1                        & 0.280                    & 0.213 (0.183,0.247)     & 0.066 (0.050,0.088)     & 0.014                    & 1\%                     & 99\%                    \\ \hline
NV                      & 6                        & 1.579                    & 0.791 (0.600,1.028)     & 0.788 (0.603,1.013)     & 0.040                    & 51\%                    & 49\%                    \\ \hline
NH                      & 4                        & 0.828                    & 0.405 (0.367,0.445)     & 0.423 (0.364,0.489)     & 0.065                    & 25\%                    & 75\%                    \\ \hline
NJ                      & 14                       & 4.797                    & 2.268 (1.961,2.616)     & 2.529 (2.117,3.009)     & 0.109                    & 4\%                     & 96\%                    \\ \hline
NM                      & 5                        & 0.966                    & 0.460 (0.394,0.535)     & 0.507 (0.410,0.617)     & 0.084                    & 14\%                    & 86\%                    \\ \hline
NY                      & 28                       & 9.021                    & 4.190 (3.467,5.027)     & 4.832 (4.175,5.567)     & 0.134                    & 1\%                     & 99\%                    \\ \hline
NC                      & 16                       & 6.186                    & 3.109 (2.562,3.769)     & 3.078 (2.358,3.963)     & 0.026                    & 59\%                    & 41\%                    \\ \hline
ND                      & 3                        & 0.371                    & 0.250 (0.203,0.305)     & 0.121 (0.091,0.159)     & 0.000                    & 100\%                   & 0\%                     \\ \hline
OH                      & 17                       & 5.974                    & 3.219 (2.862,3.632)     & 2.755 (2.399,3.195)     & 0.012                    & 92\%                    & 9\%                     \\ \hline
OK                      & 7                        & 1.564                    & 1.044 (0.960,1.135)     & 0.520 (0.449,0.607)     & 0.000                    & 100\%                   & 0\%                     \\ \hline
OR                      & 8                        & 2.478                    & 1.174 (0.982,1.397)     & 1.304 (1.048,1.596)     & 0.121                    & 7\%                     & 93\%                    \\ \hline
PA                      & 19                       & 7.218                    & 3.591 (3.046,4.221)     & 3.627 (3.185,4.118)     & 0.007                    & 44\%                    & 56\%                    \\ \hline
RI                      & 4                        & 0.534                    & 0.247 (0.215,0.283)     & 0.287 (0.248,0.330)     & 0.115                    & 3\%                     & 97\%                    \\ \hline
SC                      & 9                        & 2.801                    & 1.450 (1.168,1.790)     & 1.351 (1.063,1.702)     & 0.058                    & 96\%                    & 4\%                     \\ \hline
SD                      & 3                        & 0.428                    & 0.251 (0.213,0.294)     & 0.177 (0.145,0.219)     & 0.074                    & 99\%                    & 1\%                     \\ \hline
TN                      & 11                       & 3.235                    & 2.017 (1.692,2.404)     & 1.217 (1.020,1.449)     & 0.000                    & 100\%                   & 0\%                     \\ \hline
TX                      & 40                       & 12.701                   & 6.583 (5.473,7.874)     & 6.118 (4.592,8.092)     & 0.033                    & 87\%                    & 13\%                    \\ \hline
UT                      & 6                        & 1.725                    & 0.985 (0.746,1.273)     & 0.740 (0.519,1.020)     & 0.000                    & 97\%                    & 3\%                     \\ \hline
VT                      & 3                        & 0.378                    & 0.112 (0.091,0.138)     & 0.265 (0.216,0.322)     & 0.000                    & 0\%                     & 100\%                   \\ \hline
VA                      & 13                       & 4.852                    & 2.335 (2.106,2.616)     & 2.517 (1.975,3.202)     & 0.101                    & 18\%                    & 82\%                    \\ \hline
WA                      & 12                       & 4.368                    & 2.052 (1.715,2.446)     & 2.317 (1.821,2.899)     & 0.136                    & 2\%                     & 98\%                    \\ \hline
WV                      & 4                        & 0.812                    & 0.586 (0.474,0.717)     & 0.227 (0.149,0.365)     & 0.000                    & 100\%                   & 0\%                     \\ \hline
WI                      & 10                       & 3.359                    & 1.652 (1.422,1.911)     & 1.707 (1.476,1.961)     & 0.009                    & 39\%                    & 61\%                    \\ \hline
WY                      & 3                        & 0.278                    & 0.201 (0.181,0.224)     & 0.077 (0.060,0.097)     & 0.000                    & 100\%                   & 0\%                     \\ \hline
\end{tabular}
\caption{5000 Scenarios: Electoral Unit, Electoral Votes, Total Votes (in millions), Republican and Democratic votes (with 2.5 and 97.5 percentiles), $\lambda^*_\state$, and the win probabilities.}
\label{tab:sim_results}
\end{table}

\subsection{Network Flow Model}\label{sec: network model}

% Let $\mathbf{\Xi} \in \mathbb{R}^{z \times 2 \times l}$ be a matrix of $z$ scenarios corresponding to a simulation of the NM votes for both parties in every county. For each scenario $q \in [z]$, we denote by $\xi_{q,a,i}$ the number of NM votes in scenario $q$ for party $p_a$ in county $c_i$, where $b \in [2]$, and $i \in [l]$.

Let~$L \in \mathbb{N}$ be a fixed upper limit on the number of relocation movements. We restrict the set of possible relocations to pairs $\mathcal{A} \subseteq \mathcal{C} \times \mathcal{C}$ for practicality reasons. Specifically, we consider the destination county to be within a swing state and restrict relocation distances to $100$ miles. We also remove from~$\mathcal{A}$ relocation moves with data consistency issues and other operational difficulties. We refer to Appendix~\ref{appendix:data} for details and the relocation data sources for modeling.

% \bigskip 
% \bigskip 

% where the destination county is located in a swing state

% For the sake of computational efficiency, 

% we restrict the set of possible relocations to pairs where the destination county is located in a swing state.Moreover, for practicality reasons, we also limit relocation distances to 100 miles. Therefore, with $\hat\counties \subseteq \counties$ representing the set of counties located in swing states, we define the set~$\mathcal{X}$ of admissible pairs of origins and destinations for the relocations as 
% \begin{equation*}
%     \mathcal{X} \coloneqq \bigl\{ (i,j) \ | \ \county_i \in \counties \setminus \hat\counties; \ \county_j \in \hat\counties; \ \county_i \text{ and } \county_j \text{ are located at most 100 miles apart} \bigr\}.
% \end{equation*} 

We identify an optimal relocation strategy using the following network flow model:
% Details about relocation data are presented in Appendix~\ref{appendix:data}.
%For ease of notation and with a slight abuse of notation, we let $\mathcal{U}, \mathcal{S}$, and $\mathcal{D}$ refer both to the set of units, states, counties, and districts, respectively, and also the index set of those objects, with $u,s,$ and $d$ indexing them.  We do the same for counties where $\mathcal{C}$ will be both the set of counties and the indices set of those counties, with $i$ and $j$ as indices.
\begin{subequations}
\label{eq:flowmodel}
\begin{align}
\max_{\mathbf{x}, \mathbf{y}, \mathbf{w}, \hat{\mathbf{w}}} \quad &  \frac{1}{z}\sum_{q \in [z]} \hat{w}_{q} \label{eq:2a}
\\
\text{s.t.} \quad 
    & y_{s,q} =
\sum_{\county_i \in \mathcal{C}_s} \left( \xi_{q,1,\county_i} + \sum_{\county_j \in \mathcal{C} \setminus \mathcal{C}_s} \left( 
x_{j,i} -  x_{i,j}
\right) \right)
    && \forall s \in \mathcal{S}, \forall q \in [z] \label{eq:2b} \\ 
\quad & y_{d,q} = \sum_{\county_i \in \mathcal{C}_d} f_{\county_i,d}\left( \xi_{q,1,\county_i} + \sum_{\county_j \in \mathcal{C} \setminus \mathcal{C}_d} \left( x_{j,i} -  x_{i,j}
\right) \right) && \forall d \in \mathcal{D}, \forall q \in [z]  \label{eq:2c} \\ 
\quad & y_{u,q} +M(1-w_{u,q}) \geq \sum_{\county_i \in \mathcal{C}_u} \xi_{q,2,\county_i} + w_{u,q} 
    && \forall \unit \in \units, \forall q \in [z] \label{eq:2d} \\
\quad & \sum_{\unit \in \units} v_\unit w_{u,q} \geq 270\hat{w}_q && \forall q \in [z] \label{eq:2e}\\
\quad & \sum_{(i,j) \in \mathcal{A}} x_{i,j} \leq L \label{eq:2f} \\
\quad & 0 \leq x_{i,j} \leq \min_{q \in [z]}\{\xi_{q,1,\county_i}\} && \forall (i,j) \in \mathcal{A} \label{eq:2g} \\
% \quad & x_{i,j} = 0 && \forall (i,j) \not\in \mathcal{A}  \label{eq:2h} \\
\quad & y_{u,q} \geq 0, \ w_{u,q} \in \{0,1\} && \forall \unit \in \units, \forall q \in [z]  \label{eq:2h} \nonumber 
 \\
 \quad & \hat{w}_{q} \in \{0,1\} && \forall q \in [z]. \nonumber
\end{align}
\label{eq:networkFlow}
\end{subequations}

Variable $y_{u,q}$ in \eqref{eq:flowmodel} represents the total number of votes for party $\party_1$ in unit $\unit$ under the $q$-th scenario after the movements. Binary variable $w_{u,q}$ equals 1 if and only if $\party_1$ wins the popular vote in unit $\unit$ under $q$. Similarly, binary variable $\hat{w}_{q}$ indicates if $\party_1$ wins the election in $q$. Each variable $x_{i,j}$ represents the number of people moved from county $\county_i$ to county $\county_j$. 

% For the sake of computational efficiency, we restrict the set of possible relocations to pairs where the destination county is located in a swing state.Moreover, for practicality reasons, we also limit relocation distances to 100 miles. Therefore, with $\hat\counties \subseteq \counties$ representing the set of counties located in swing states, we define the set~$\mathcal{A}$ of admissible pairs of origins and destinations for the relocations as 
% \begin{equation*}
%     \mathcal{X} \coloneqq \bigl\{ (i,j) \ | \ \county_i \in \counties \setminus \hat\counties; \ \county_j \in \hat\counties; \ \county_i \text{ and } \county_j \text{ are located at most 100 miles apart} \bigr\}.
% \end{equation*} 
% We also remove from~$\mathcal{X}$ relocation moves with data consistency issues and other technical complications (see Appendix~\ref{appendix:data}). V

Our goal in~\eqref{eq:2a} is to maximize the estimated win probability of~$\party_1$. Constraints~\eqref{eq:2b}  and \eqref{eq:2c} count the votes for each state and district, respectively, per scenario after the relocations; recall that~$f_{\county_i,d}$, used in~\eqref{eq:2c}, is the fraction of county $\county_i$ within district unit $d$. Constraints~\eqref{eq:2d} and~\eqref{eq:2e} set the activation variables indicating the victory per state and nationwide, respectively. Constraint~\eqref{eq:2f} limits the total number of movements to~$L$ voters. Lastly, constraints~\eqref{eq:2g} limit the number of movements out of each county to the minimum number of voters observed in that county across all scenarios.

\smallskip
\noindent \textbf{Post-optimization Adjustments.}
We implement two post-optimization routines to fine-tune the solutions obtained from the network flow model. First, we avoid excessively small relocation sizes by transferring movements between pairs $(i,j)$ involving fewer than $1,000$ people to other movements with more than $1,000$ people whenever possible.
% replacing movements between pairs~$(i,j)$ involving less than $1,000$ people by redirecting the same number of movements to another pair~$(i',j')$ that is already present in the solution (moving more that $1,000$ people).
%The first routine seeks to avoid excessively small movements between counties, as follows. If there are any movements between counties of less than $1,000$ people, our routine attempts to aggregate this movement with another one already present in the solution. 
If the resulting aggregation violates any upper bound constraints, our post-optimization routine simply reverts back to the original solution.

% We implement two post-optimization routines to fine-tune the solutions obtained from the network flow model. First, we avoid excessively small relocation sizes by replacing movements between pairs~$(i,j)$ involving less than $1,000$ people by redirecting the same number of movements to another pair~$(i',j')$ that is already present in the solution (moving more that $1,000$ people).
% %The first routine seeks to avoid excessively small movements between counties, as follows. If there are any movements between counties of less than $1,000$ people, our routine attempts to aggregate this movement with another one already present in the solution. 
% If the resulting aggregation violates any upper bound constraints, our post-optimization routine simply reverts back to the original solution. 

A second post-optimization routine tries to reduce the total distance traversed by all movements
%. This aspect is not handled by the network model, so 
%Since any movement within 100 miles of distance is feasible in our original solution, there is no incentive to favour smaller distances over longer ones. Thus, 
%our second post-optimization routine seeks to minimize the total distance traversed by all movements 
while maintaining the same aggregate flows out of each state and the same aggregate flows into each state as in the original solution. Note that such modifications do not change the estimated win probability.  

% \subsubsection{Training and Testing Data}
\smallskip
\noindent \textbf{Training and Testing Data.}
A collection of scenario batches were used for model calibration and testing. For the results reported below, we considered a new training set of 1,000 random scenarios (not used before for calibration or testing) as input to our optimization models. After obtaining an optimal relocation plan, we considered another test set of 5,000 scenarios  
to compute win probabilities and supporting statistics.

\section{Small Movements, Large Impact}

The optimization model identifies relocation plans that maximize win probabilities by selecting 10,000, 15,000, 20,000, 25,000, 50,000, 100,000, 150,000, 200,000, and 250,000 voters in favor of each candidate, solved separately for each such level. All optimization models were solved using a standard commercial mixed-integer linear programming solver (CPLEX 20.1) to optimality in less than one hour. 

\begin{figure}[h!]
    \centering
    \includegraphics[width=0.6\textwidth]{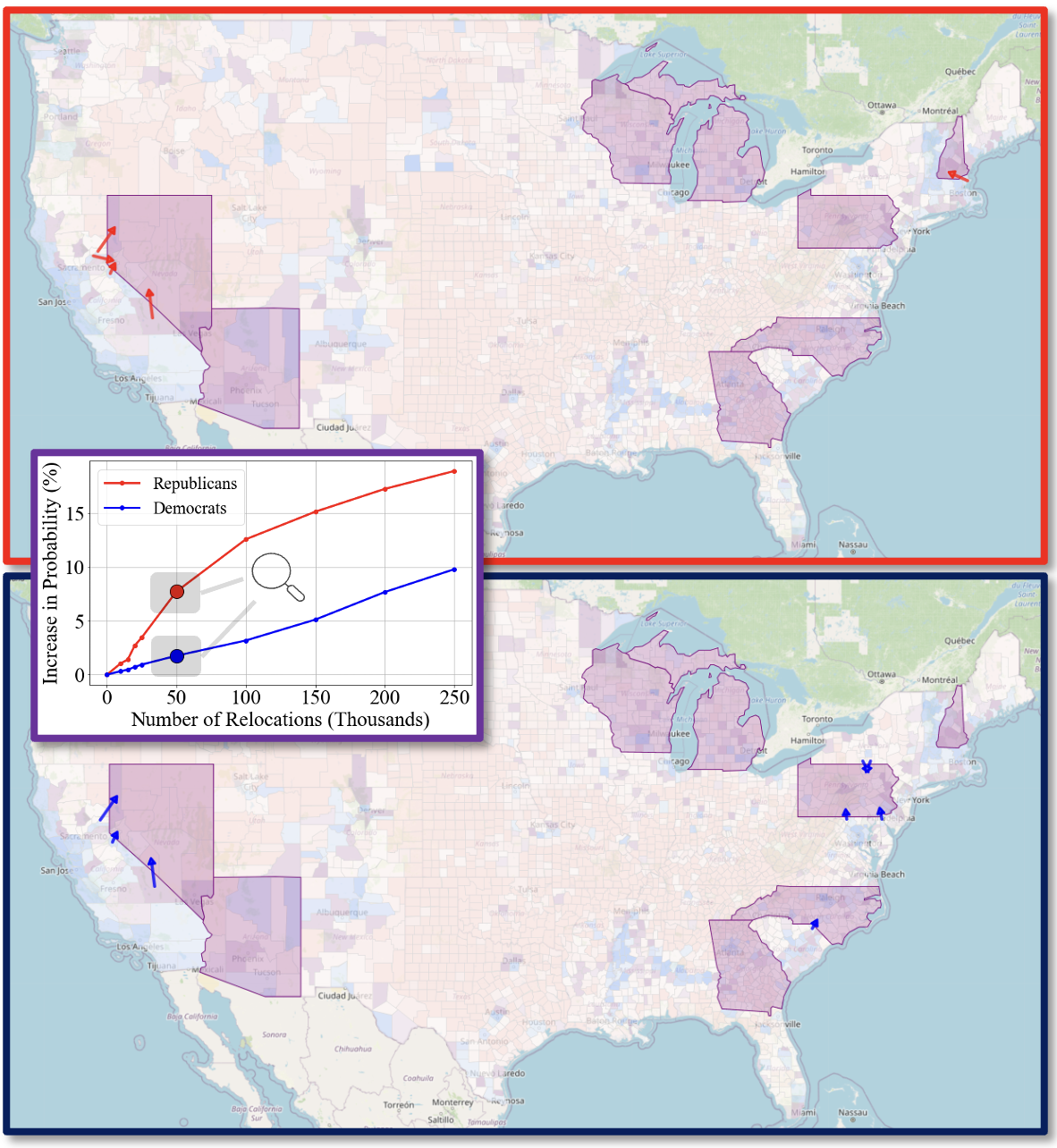} % Replace with your figure's path
    \caption{Movement patterns for 50,000 people for Republicans (top) and Democrats (bottom), and impact of the number of people moved on the probability of winning for different relocation sizes (line plot).}
    \label{fig:movement-probability}
\end{figure}

The line plot in Figure~\ref{fig:movement-probability}  shows the increase in win probability that would be realized for each candidate as a function of the number of people moving.  Surprisingly, for 10,000 moves, the Republican candidate could already realize a 1.06\% increase in election win probability, and with 250,000 moves, an 18.92\% increase.  For the Democratic candidate, changes are still substantial but less effective: 10,000 and 250,000 moves increase the Democratic win probability by 0.34\% and 9.8\%, respectively.

To  explain the cause of this disparity between the candidates for the current election, the maps in Figure~\ref{fig:movement-probability} depict the prescribed relocation strategies for 50,000 people, 
% which reveals substantial differences in the effect of optimal relocation strategies 
for both the Republican (top) and the Democratic (bottom) candidates. We observe that 7.72\% and 1.76\% win probability increases could be attained for the Republican and Democratic candidates in this setting, respectively. For the Republican candidate, the model suggests large movements into New Hampshire (34,755) and Nevada (15,244).  For the Democratic candidate, the focus is on Pennsylvania (41,754), accompanied by relatively few moves into Nevada (4,072) and North Carolina (4,173).
%All county-to-county prescribed movements are detailed in the SM (tables~\ref{tab:R_50000} and~\ref{tab:D_50000}). 
In this scenario, the average weighted distances traveled by Republican and Democratic supporters are a mere 57.31 and 33.94 miles, respectively.  Detailed results for all relocation sizes tested are provided in Appendix~\ref{appendix:movements}. 

We hypothesize that the difference in relocation strategy efficiency between the two parties is due to New Hampshire's small size and current political lean. With only four electoral votes and about 800,000 registered voters, a significant shift in New Hampshire's outcome can be achieved with just a few relocations. Specifically, relocating 35,000 Republican voters would increase their candidate's win probability from 24.71\% to 73.02\%, and considering a total number of movements of 50,000, this still allows additional moves into Nevada, boosting its win probability. In this plan, Republican voters are moving only from Massachusetts and California, states where the Republican candidate is unlikely to win.  In contrast, Democrats focus on Pennsylvania, where the candidate has a 55.58\% chance of winning. Pennsylvania's larger population of 8.8 million registered voters requires more relocations -- 42,000 voters only increase the Democratic candidate's win probability to 61.62\%. The remaining 8,000 relocations are insufficient to significantly impact other states, unlike the Republican's dual strategy with New Hampshire and Nevada.

These results are surprising for two reasons. First,  they suggest that engaging less than 0.02\% of the U.S. voting population is sufficient to increase the Republican candidate's chance of winning by almost 8\%. Second, with the same number of movements, the increase in the Democratic candidate's chances of winning do not exceed 2\%. 
% The changes in the win probabilities are not as pronounced for the Democratic candidate, but remain substantial compared to the required fraction of eligible voters. 
Asymmetries like this are
 %c partisan impact is 
 not uncommon; for example, rain and snow hurt turnout, and these fluctuations have historically been more beneficial for Republicans~\citep{gomez2007republicans}.

 Note that the long-term impact of the movement patterns suggested  is complex to estimate. The 50,000-person movement for Republican voters, for example, involves relocating individuals from Democratic strongholds to purple states, potentially achieving a dual effect: flipping a purple state red and increasing the Republican electoral vote count (after a new census) in future elections.

% Moreover, the results reveal the arguably unexpected key role that states with small populations can have in the U.S. elections. For the 2024 elections, New Hampshire could be a pivot battlefield if enough relocation transpired.

\section{Legal, Financial, and Logistical Considerations}

How feasible would such a relocation strategy be in practice?
A close inspection of the main potential hurdles indicates that it would be far less challenging to implement than what common perception might suggest.

Current U.S. legislation does not impose barriers on relocation strategies. There is a 30-day limit on residency requirements for voting in federal elections, including presidential elections, based on the 1970 amendments to the Voting Rights Act (42 U.S.C. § 1973aa-1). This federal law stipulates that no state may impose a residency requirement of more than 30 days for voting in any presidential election.  Additionally, the U.S. Supreme Court reinforced this in the 1972 case Dunn v. Blumstein (405 U.S. 330), where the Court ruled that lengthy residency requirements violated the Equal Protection Clause of the 14$^{\textnormal{th}}$ Amendment, despite what some politicians might have called for \citep{ImpInterstate}. Therefore, relocation translates to voting eligibility in a relatively short period. 
% In the U.S., public debate on the impact of moves in the electoral process has motivated the discussion of policies preventing inter-state movers from voting for a few years.
%Recently, an U.S. Representative has expressed concerns with the impact of moves in the electoral process, and advocated for policies that would prevent people from voting for a few years after moving to a different state

Another consideration is engagement, as the relocation burden for a voter would be far greater than just voting in their current state of residence. Although turnout initiatives typically focus on motivating the less engaged segments of voters, the small number of movements observed in our study suggest that 
% bad actors could focus solely on 
attention should be placed on enthusiastic voters who are already planning to vote.  Election rallies frequently have tens of thousands of people in attendance~\citep{peoples2024}.
%and the most recent Republican National Convention was attended by 50,000 people.
Considering that the average moving distance in our results is less than 60 miles, such a relocation may have little impact on an individual's social and professional life.  Note that 50,000 interstate relocations are far less than what typically happens annually in the U.S., where nearly 8 million people have moved between states in recent years in an un-coordinated fashion~\citep{8MMovers}. Moreover, the 100 companies with the largest workforce in the U.S. employ more than 80,000 people each~\citep{workforce2024}, so even a corporate relocation could move the needle and be decisive in any given election.

From a financial standpoint, an estimate for the total expense required to move up to 50,000 people is far less than the amount spent by campaigns during each election. For example, as of August 15, 2024, the estimated cost of moving a two-bedroom house from Lowell, MA to Nashua, NH is between \$2,081 and \$2,542~\citep{moving2024}. Using the midpoint of this cost range, a rough estimate for the total cost of moving 50,000 people is around \$150 million. For reference, over 200,000 people have donated over \$3,300 to 2024 presidential election campaigns by August 15, 2024~\citep{opensecrets2024}. Instead of donating to a campaign, the money could be used to move passionate voters and directly impact vote totals, noting that campaign dollars do not necessarily translate to votes \citep{BadAds}.

%Moreover, these towns are less than 30 miles apart, so such a relocation would hardly affect someone's employment and social life.
%still enable someone to maintain their employment and proximity to family and friends. 
% To put this in perspective, 
%we notice that the electoral expenses with advertisement  of both parties combinedwere superior to U\$ 1.3 Billion in the 2020 elections\footnote{ \url{https://www.opensecrets.org/2020-presidential-race}}; in particular, the U\$450 Million spent in online advertisements alone would be sufficient to move approximately 100,000 people that distance.
%Electoral campaigns are mostly focused on voter persuasion and turnout. The 
% advertisement expenses for both parties combined surpassed \$1.3 Billion in the 2020 election~\cite{opensecrets2020}.  Additionally, 
% %\footnote{ \url{https://www.opensecrets.org/2020-presidential-race}}; 
% according to the U.S.\ Federal Election Commission~\cite{washburn2013}, the total cost of presidential campaigns has been increasing at about 10\% per year over the last several decades, considerably outpacing inflation.
% Our study suggests that with much less money, if applied to a coordinated movement of voters, would be more effective than advertisements, despite targeting significantly fewer voters.

% into voting outcomes through ads is challenging, and may even have negative effects, whereas a move by a passionate voter would unquestionably impact the vote count.  

% Understanding the reasons and voting implications of migration flows is challenging and has gained attention worldwide and domestically~\cite{CHOU_DANCYGIER_2021,2024movementVoters}.  

\section{Conclusion}

In this paper, we explore whether a small but strategic movement of people could meaningfully influence the outcome of a U.S. presidential election. Specifically, we approach this question through an analytical lens, leveraging predictive models and operations research methodologies. Although this idea has often been dismissed as impractical, politicians have, in fact, called on people to relocate to vote in key elections \citep{yang}. By building on the structure of the electoral college and examining the proximity of swing states to neighboring states with solid partisan majorities, we demonstrate that even modest population shifts can significantly alter win probabilities.

However, these movements would require careful coordination. One could envision a grassroots effort where motivated voters from critical border counties unite to push their preferred candidate over the finish line. Even more alarming is the prospect of well-funded, malicious actors paying voters to relocate, amplifying their political power in a targeted way. It is important to clarify that the authors neither endorse nor condone these relocation strategies; instead, we aim to shed light on how such planned efforts could be exploited and their potentially profound impact on the democratic process.

Whether or not these tactics should be considered a violation of voting laws, or merit new legislative measures to detect or prevent such collective actions, is beyond the scope of this paper. Nonetheless, it is an issue that warrants further policy discussion and debate. We hope this work will inspire future research in operations modeling and algorithms, contributing to more resilient and equitable election systems.

\bibliographystyle{plainnat}
\bibliography{references.bib}

\newpage 

% \begin{APPENDICES}
% \begin{APPENDIX}
\appendix

\section{Relocation Data}\label{appendix:data}

 Movements in and out were prohibited for a few states, either for simplicity or due to data challenges. Specifically, a) due to distance, we do not allow movements in or out of Alaska and Hawaii; b) 
 due to district electoral units, we do not allow movements in or out of Maine and Nebraska; and c) due to discrepancies in naming conventions, movements in and out of Connecticut were also prohibited. In the voting data for Connecticut, the voting records are broken down by Planning Regions, as opposed to CT Counties (see e.g. \url{https://portal.ct.gov/csl/research/ct-towns-counties?language=en_US}. A similar discrepancy is also present in Alaska, noting that  movements are restricted  in the model anyways.

The distance between two counties is calculated as the distance between the centroid of their cities weighted by the population of such cities. Specifically, for every city, we first extract the population, latitude, longitude, and county from SimpleMaps (\url{https://simplemaps.com/}). Next, we compute the coordinates of the county centroids as the weighted average of the latitude and longitude of the cities in each county, weighted by population size.

% we consider the following procedure: 
% \begin{itemize}
%     \item For every city, we extract the population, latitude, longitude, and county from SimpleMaps (\url{https://simplemaps.com/});
%     \item The coordinates of the county centroids are calculated as the weighted average of the latitude and longitude of the cities in each county, weighted by population.
% \end{itemize} 

%  \begin{itemize}
%     \item Due to distance, we do not allow movements in or out of Alaska and Hawaii;
%     \item Due to district electoral units, we do not allow movements in or out of Maine and Nebraska;
%     \item Due to discrepancies in naming conventions, movements in and out of Connecticut were also prohibited.   In the voting data for Connecticut, the voting records are broken down by Planning Regions, as opposed to CT Counties (see e.g. \url{https://portal.ct.gov/csl/research/ct-towns-counties?language=en_US}.
%    % \citeAppendix{CTBreakdown}.  
%     A similar discrepancy also presents in Alaska, noting that  movements are restricted  in the model anyways.
% \end{itemize}   

\section{Movement Patterns}\label{appendix:movements}

 % \begin{APPENDICES}

% \section{Appendix}

This section presents tables with detailed movement patterns  for each of the two candidates, for the number of movements tested (10,000, 15,000, 20,000, 25,000, 50,000, 100,000, 150,000, 200,000, and 250,000 people).  Each table presents the origin county and state, the destination county and state, the movement (number of people), and the distance (in miles) between the counties. The tables also show the movements aggregated at the state level.

\begin{table}[H]
\begin{center}
\begin{scriptsize}
\begin{tabular}{|c|c|c|c|c|c|}
\hline
\multicolumn{2}{|c|}{\textbf{Origin}} & \multicolumn{2}{c|}{\textbf{Destination}} & {\centering \textbf{Movement}} & {\centering \textbf{Distance}} \\ \cline{1-4}
\textbf{County} & \textbf{State} & \textbf{County} & \textbf{State} &  &  \\ \hline
& & & \textbf{New Hampshire} & \textbf{10000} & \\
\cdashline{1-6}
Essex & Massachusetts & Hillsborough & New Hampshire & 10000 & 45 \\
\hline
\end{tabular}
\end{scriptsize}
\end{center}
\caption{Republicans (10,000 movements)}
\label{tab:R_10000}\end{table}
% \vspace{0.5cm}

\begin{table}[H]
\begin{center}
\begin{scriptsize}
\begin{tabular}{|c|c|c|c|c|c|}
\hline
\multicolumn{2}{|c|}{\textbf{Origin}} & \multicolumn{2}{c|}{\textbf{Destination}} & {\centering \textbf{Movement}} & {\centering \textbf{Distance}} \\ \cline{1-4}
\textbf{County} & \textbf{State} & \textbf{County} & \textbf{State} &  &  \\ \hline
& & & \textbf{Georgia} & \textbf{1039} & \\
\cdashline{1-6}
Nassau & Florida & Camden & Georgia & 1039 & 19 \\
\hline
& & & \textbf{Nevada} & \textbf{4787} & \\
\cdashline{1-6}
Inyo & California & Esmeralda & Nevada & 3344 & 78 \\
Sierra & California & Washoe & Nevada & 555 & 47 \\
Nevada & California & Carson & Nevada & 543 & 90 \\
Alpine & California & Douglas & Nevada & 343 & 26 \\
\hline
& & & \textbf{North Carolina} & \textbf{4173} & \\
\cdashline{1-6}
Marlboro & South Carolina & Scotland & North Carolina & 4173 & 21 \\
\hline
\end{tabular}
\end{scriptsize}
\end{center}
\caption{Democrats (10,000 movements)}
\label{tab:D_10000}\end{table}
% \vspace{0.5cm}

\begin{table}[H]
\begin{center}
\begin{scriptsize}
\begin{tabular}{|c|c|c|c|c|c|}
\hline
\multicolumn{2}{|c|}{\textbf{Origin}} & \multicolumn{2}{c|}{\textbf{Destination}} & {\centering \textbf{Movement}} & {\centering \textbf{Distance}} \\ \cline{1-4}
\textbf{County} & \textbf{State} & \textbf{County} & \textbf{State} &  &  \\ \hline
& & & \textbf{Nevada} & \textbf{7005} & \\
\cdashline{1-6}
Inyo & California & Esmeralda & Nevada & 4056 & 78 \\
Nevada & California & Carson & Nevada & 1963 & 90 \\
Sierra & California & Washoe & Nevada & 854 & 47 \\
Alpine & California & Douglas & Nevada & 130 & 26 \\
\hline
& & & \textbf{New Hampshire} & \textbf{7994} & \\
\cdashline{1-6}
Essex & Massachusetts & Hillsborough & New Hampshire & 7994 & 45 \\
\hline
\end{tabular}
\end{scriptsize}
\end{center}
\caption{Republicans (15,000 movements)}
\label{tab:R_15000}\end{table}
% \vspace{0.5cm}

\begin{table}[H]
\begin{center}
\begin{scriptsize}
\begin{tabular}{|c|c|c|c|c|c|}
\hline
\multicolumn{2}{|c|}{\textbf{Origin}} & \multicolumn{2}{c|}{\textbf{Destination}} & {\centering \textbf{Movement}} & {\centering \textbf{Distance}} \\ \cline{1-4}
\textbf{County} & \textbf{State} & \textbf{County} & \textbf{State} &  &  \\ \hline
& & & \textbf{Georgia} & \textbf{1039} & \\
\cdashline{1-6}
Nassau & Florida & Camden & Georgia & 1039 & 19 \\
\hline
& & & \textbf{Nevada} & \textbf{4942} & \\
\cdashline{1-6}
Inyo & California & Esmeralda & Nevada & 3344 & 78 \\
Nevada & California & Carson & Nevada & 698 & 90 \\
Sierra & California & Washoe & Nevada & 555 & 47 \\
Alpine & California & Douglas & Nevada & 343 & 26 \\
\hline
& & & \textbf{New Hampshire} & \textbf{4844} & \\
\cdashline{1-6}
Washington & New York & Sullivan & New Hampshire & 4844 & 99 \\
\hline
& & & \textbf{North Carolina} & \textbf{4173} & \\
\cdashline{1-6}
Marlboro & South Carolina & Scotland & North Carolina & 4173 & 21 \\
\hline
\end{tabular}
\end{scriptsize}
\end{center}
\caption{Democrats (15,000 movements)}
\label{tab:D_15000}\end{table}
% \vspace{0.5cm}

\begin{table}[H]
\begin{center}
\begin{scriptsize}
\begin{tabular}{|c|c|c|c|c|c|}
\hline
\multicolumn{2}{|c|}{\textbf{Origin}} & \multicolumn{2}{c|}{\textbf{Destination}} & {\centering \textbf{Movement}} & {\centering \textbf{Distance}} \\ \cline{1-4}
\textbf{County} & \textbf{State} & \textbf{County} & \textbf{State} &  &  \\ \hline
& & & \textbf{New Hampshire} & \textbf{20000} & \\
\cdashline{1-6}
Essex & Massachusetts & Hillsborough & New Hampshire & 20000 & 45 \\
\hline
\end{tabular}
\end{scriptsize}
\end{center}
\caption{Republicans (20,000 movements)}
\label{tab:R_20000}\end{table}
% \vspace{0.5cm}

\begin{table}[H]
\begin{center}
\begin{scriptsize}
\begin{tabular}{|c|c|c|c|c|c|}
\hline
\multicolumn{2}{|c|}{\textbf{Origin}} & \multicolumn{2}{c|}{\textbf{Destination}} & {\centering \textbf{Movement}} & {\centering \textbf{Distance}} \\ \cline{1-4}
\textbf{County} & \textbf{State} & \textbf{County} & \textbf{State} &  &  \\ \hline
& & & \textbf{Georgia} & \textbf{1278} & \\
\cdashline{1-6}
Mccormick & South Carolina & Lincoln & Georgia & 1278 & 19 \\
\hline
& & & \textbf{Nevada} & \textbf{4072} & \\
\cdashline{1-6}
Inyo & California & Esmeralda & Nevada & 3172 & 78 \\
Sierra & California & Washoe & Nevada & 555 & 47 \\
Alpine & California & Douglas & Nevada & 343 & 26 \\
\hline
& & & \textbf{North Carolina} & \textbf{4173} & \\
\cdashline{1-6}
Marlboro & South Carolina & Scotland & North Carolina & 4173 & 21 \\
\hline
& & & \textbf{Pennsylvania} & \textbf{10475} & \\
\cdashline{1-6}
New Castle & Delaware & Chester & Pennsylvania & 5237 & 36 \\
Camden & New Jersey & Philadelphia & Pennsylvania & 5237 & 18 \\
\hline
\end{tabular}
\end{scriptsize}
\end{center}
\caption{Democrats (20,000 movements)}
\label{tab:D_20000}\end{table}
% \vspace{0.5cm}

\begin{table}[H]
\begin{center}
\begin{scriptsize}
\begin{tabular}{|c|c|c|c|c|c|}
\hline
\multicolumn{2}{|c|}{\textbf{Origin}} & \multicolumn{2}{c|}{\textbf{Destination}} & {\centering \textbf{Movement}} & {\centering \textbf{Distance}} \\ \cline{1-4}
\textbf{County} & \textbf{State} & \textbf{County} & \textbf{State} &  &  \\ \hline
& & & \textbf{New Hampshire} & \textbf{25000} & \\
\cdashline{1-6}
Essex & Massachusetts & Hillsborough & New Hampshire & 25000 & 45 \\
\hline
\end{tabular}
\end{scriptsize}
\end{center}
\caption{Republicans (25,000 movements)}
\label{tab:R_25000}\end{table}
% \vspace{0.5cm}

\begin{table}[H]
\begin{center}
\begin{scriptsize}
\begin{tabular}{|c|c|c|c|c|c|}
\hline
\multicolumn{2}{|c|}{\textbf{Origin}} & \multicolumn{2}{c|}{\textbf{Destination}} & {\centering \textbf{Movement}} & {\centering \textbf{Distance}} \\ \cline{1-4}
\textbf{County} & \textbf{State} & \textbf{County} & \textbf{State} &  &  \\ \hline
& & & \textbf{Georgia} & \textbf{1710} & \\
\cdashline{1-6}
Nassau & Florida & Camden & Georgia & 1710 & 19 \\
\hline
& & & \textbf{Nevada} & \textbf{4631} & \\
\cdashline{1-6}
Inyo & California & Esmeralda & Nevada & 3344 & 78 \\
Sierra & California & Washoe & Nevada & 555 & 47 \\
Nevada & California & Carson & Nevada & 387 & 90 \\
Alpine & California & Douglas & Nevada & 343 & 26 \\
\hline
& & & \textbf{North Carolina} & \textbf{4173} & \\
\cdashline{1-6}
Marlboro & South Carolina & Scotland & North Carolina & 4173 & 21 \\
\hline
& & & \textbf{Pennsylvania} & \textbf{14484} & \\
\cdashline{1-6}
New Castle & Delaware & Chester & Pennsylvania & 14484 & 36 \\
\hline
\end{tabular}
\end{scriptsize}
\end{center}
\caption{Democrats (25,000 movements)}
\label{tab:D_25000}\end{table}
% \vspace{0.5cm}

\begin{table}[H]
\begin{center}
\begin{scriptsize}
\begin{tabular}{|c|c|c|c|c|c|}
\hline
\multicolumn{2}{|c|}{\textbf{Origin}} & \multicolumn{2}{c|}{\textbf{Destination}} & {\centering \textbf{Movement}} & {\centering \textbf{Distance}} \\ \cline{1-4}
\textbf{County} & \textbf{State} & \textbf{County} & \textbf{State} &  &  \\ \hline
& & & \textbf{Nevada} & \textbf{15244} & \\
\cdashline{1-6}
Nevada & California & Carson & Nevada & 10202 & 90 \\
Inyo & California & Esmeralda & Nevada & 4056 & 78 \\
Sierra & California & Washoe & Nevada & 854 & 47 \\
Alpine & California & Douglas & Nevada & 130 & 26 \\
\hline
& & & \textbf{New Hampshire} & \textbf{34755} & \\
\cdashline{1-6}
Essex & Massachusetts & Hillsborough & New Hampshire & 34755 & 45 \\
\hline
\end{tabular}
\end{scriptsize}
\end{center}
\caption{Republicans (50,000 movements)}
\label{tab:R_50000}\end{table}
% \vspace{0.5cm}

\begin{table}[H]
\begin{center}
\begin{scriptsize}
\begin{tabular}{|c|c|c|c|c|c|}
\hline
\multicolumn{2}{|c|}{\textbf{Origin}} & \multicolumn{2}{c|}{\textbf{Destination}} & {\centering \textbf{Movement}} & {\centering \textbf{Distance}} \\ \cline{1-4}
\textbf{County} & \textbf{State} & \textbf{County} & \textbf{State} &  &  \\ \hline
& & & \textbf{Nevada} & \textbf{4072} & \\
\cdashline{1-6}
Inyo & California & Esmeralda & Nevada & 3172 & 78 \\
Sierra & California & Washoe & Nevada & 555 & 47 \\
Alpine & California & Douglas & Nevada & 343 & 26 \\
\hline
& & & \textbf{North Carolina} & \textbf{4173} & \\
\cdashline{1-6}
Marlboro & South Carolina & Scotland & North Carolina & 4173 & 21 \\
\hline
& & & \textbf{Pennsylvania} & \textbf{41754} & \\
\cdashline{1-6}
New Castle & Delaware & Chester & Pennsylvania & 19741 & 36 \\
Washington & Maryland & Franklin & Pennsylvania & 11006 & 24 \\
Tioga & New York & Bradford & Pennsylvania & 5638 & 28 \\
Chemung & New York & Bradford & Pennsylvania & 5367 & 33 \\
\hline
\end{tabular}
\end{scriptsize}
\end{center}
\caption{Democrats (50,000 movements)}
\label{tab:D_50000}\end{table}
% \vspace{0.5cm}

\begin{table}[H]
\begin{center}
\begin{scriptsize}
\begin{tabular}{|c|c|c|c|c|c|}
\hline
\multicolumn{2}{|c|}{\textbf{Origin}} & \multicolumn{2}{c|}{\textbf{Destination}} & {\centering \textbf{Movement}} & {\centering \textbf{Distance}} \\ \cline{1-4}
\textbf{County} & \textbf{State} & \textbf{County} & \textbf{State} &  &  \\ \hline
& & & \textbf{Arizona} & \textbf{12473} & \\
\cdashline{1-6}
Imperial & California & Yuma & Arizona & 12473 & 96 \\
\hline
& & & \textbf{Georgia} & \textbf{1661} & \\
\cdashline{1-6}
Hamilton & Tennessee & Catoosa & Georgia & 1661 & 15 \\
\hline
& & & \textbf{Nevada} & \textbf{29885} & \\
\cdashline{1-6}
Nevada & California & Carson & Nevada & 19070 & 90 \\
El Dorado & California & Douglas & Nevada & 5772 & 91 \\
Inyo & California & Esmeralda & Nevada & 4056 & 78 \\
Sierra & California & Washoe & Nevada & 854 & 47 \\
Alpine & California & Douglas & Nevada & 130 & 26 \\
\hline
& & & \textbf{New Hampshire} & \textbf{42163} & \\
\cdashline{1-6}
Essex & Massachusetts & Hillsborough & New Hampshire & 42163 & 45 \\
\hline
& & & \textbf{North Carolina} & \textbf{13816} & \\
\cdashline{1-6}
Unicoi & Tennessee & Yancey & North Carolina & 4883 & 29 \\
Johnson & Tennessee & Watauga & North Carolina & 4573 & 30 \\
Carter & Tennessee & Avery & North Carolina & 2698 & 37 \\
Marlboro & South Carolina & Scotland & North Carolina & 1661 & 21 \\
\hline
\end{tabular}
\end{scriptsize}
\end{center}
\caption{Republicans (100,000 movements)}
\label{tab:R_100000}\end{table}
% \vspace{0.5cm}
\begin{table}[H]
\begin{center}
\begin{scriptsize}
\begin{tabular}{|c|c|c|c|c|c|}
\hline
\multicolumn{2}{|c|}{\textbf{Origin}} & \multicolumn{2}{c|}{\textbf{Destination}} & {\centering \textbf{Movement}} & {\centering \textbf{Distance}} \\ \cline{1-4}
\textbf{County} & \textbf{State} & \textbf{County} & \textbf{State} &  &  \\ \hline
& & & \textbf{Nevada} & \textbf{31087} & \\
\cdashline{1-6}
Nevada & California & Carson & Nevada & 25514 & 90 \\
Inyo & California & Esmeralda & Nevada & 3344 & 78 \\
El Dorado & California & Douglas & Nevada & 1328 & 91 \\
Sierra & California & Washoe & Nevada & 555 & 47 \\
Alpine & California & Douglas & Nevada & 343 & 26 \\
\hline
& & & \textbf{New Hampshire} & \textbf{6441} & \\
\cdashline{1-6}
Washington & New York & Sullivan & New Hampshire & 6441 & 99 \\
\hline
& & & \textbf{North Carolina} & \textbf{62471} & \\
\cdashline{1-6}
York & South Carolina & Mecklenburg & North Carolina & 43757 & 33 \\
Cherokee & South Carolina & Cleveland & North Carolina & 6703 & 25 \\
Chesterfield & South Carolina & Anson & North Carolina & 6595 & 32 \\
Marlboro & South Carolina & Scotland & North Carolina & 5109 & 21 \\
Dillon & South Carolina & Robeson & North Carolina & 306 & 37 \\
\hline
\end{tabular}
\end{scriptsize}
\end{center}
\caption{Democrats (100,000 movements)}
\label{tab:D_100000}\end{table}
% \vspace{0.5cm}

\begin{table}[H]
\begin{center}
\begin{scriptsize}
\begin{tabular}{|c|c|c|c|c|c|}
\hline
\multicolumn{2}{|c|}{\textbf{Origin}} & \multicolumn{2}{c|}{\textbf{Destination}} & {\centering \textbf{Movement}} & {\centering \textbf{Distance}} \\ \cline{1-4}
\textbf{County} & \textbf{State} & \textbf{County} & \textbf{State} &  &  \\ \hline
& & & \textbf{Arizona} & \textbf{12473} & \\
\cdashline{1-6}
Imperial & California & Yuma & Arizona & 6462 & 96 \\
Mckinley & New Mexico & Apache & Arizona & 6010 & 63 \\
\hline
& & & \textbf{Nevada} & \textbf{40937} & \\
\cdashline{1-6}
Nevada & California & Carson & Nevada & 19070 & 90 \\
El Dorado & California & Douglas & Nevada & 16824 & 91 \\
Inyo & California & Esmeralda & Nevada & 4056 & 78 \\
Sierra & California & Washoe & Nevada & 854 & 47 \\
Alpine & California & Douglas & Nevada & 130 & 26 \\
\hline
& & & \textbf{New Hampshire} & \textbf{51054} & \\
\cdashline{1-6}
Essex & Massachusetts & Hillsborough & New Hampshire & 51054 & 45 \\
\hline
& & & \textbf{North Carolina} & \textbf{32949} & \\
\cdashline{1-6}
Cherokee & South Carolina & Cleveland & North Carolina & 12395 & 25 \\
Danville & Virginia & Caswell & North Carolina & 5967 & 19 \\
Grayson & Virginia & Alleghany & North Carolina & 4907 & 16 \\
Chesterfield & South Carolina & Anson & North Carolina & 4098 & 32 \\
Marlboro & South Carolina & Scotland & North Carolina & 3507 & 21 \\
Galax & Virginia & Alleghany & North Carolina & 1425 & 25 \\
Carroll & Virginia & Surry & North Carolina & 646 & 30 \\
\hline
& & & \textbf{Pennsylvania} & \textbf{2215} & \\
\cdashline{1-6}
New Castle & Delaware & Chester & Pennsylvania & 2215 & 36 \\
\hline
& & & \textbf{Wisconsin} & \textbf{10370} & \\
\cdashline{1-6}
Lake & Illinois & Kenosha & Wisconsin & 10370 & 27 \\
\hline
\end{tabular}
\end{scriptsize}
\end{center}
\caption{Republicans (150,000 movements)}
\label{tab:R_150000}\end{table}
% \vspace{0.5cm}

\begin{table}[H]
\begin{center}
\begin{scriptsize}
\begin{tabular}{|c|c|c|c|c|c|}
\hline
\multicolumn{2}{|c|}{\textbf{Origin}} & \multicolumn{2}{c|}{\textbf{Destination}} & {\centering \textbf{Movement}} & {\centering \textbf{Distance}} \\ \cline{1-4}
\textbf{County} & \textbf{State} & \textbf{County} & \textbf{State} &  &  \\ \hline
& & & \textbf{Georgia} & \textbf{10765} & \\
\cdashline{1-6}
Nassau & Florida & Camden & Georgia & 10765 & 19 \\
\hline
& & & \textbf{Michigan} & \textbf{4125} & \\
\cdashline{1-6}
Lucas & Ohio & Monroe & Michigan & 2643 & 28 \\
Lagrange & Indiana & St. Joseph & Michigan & 808 & 25 \\
Elkhart & Indiana & Cass & Michigan & 673 & 33 \\
\hline
& & & \textbf{Nevada} & \textbf{33217} & \\
\cdashline{1-6}
Nevada & California & Carson & Nevada & 25514 & 90 \\
El Dorado & California & Douglas & Nevada & 3459 & 91 \\
Inyo & California & Esmeralda & Nevada & 3344 & 78 \\
Sierra & California & Washoe & Nevada & 555 & 47 \\
Alpine & California & Douglas & Nevada & 343 & 26 \\
\hline
& & & \textbf{New Hampshire} & \textbf{8883} & \\
\cdashline{1-6}
Essex & Massachusetts & Hillsborough & New Hampshire & 8883 & 45 \\
\hline
& & & \textbf{North Carolina} & \textbf{93008} & \\
\cdashline{1-6}
York & South Carolina & Mecklenburg & North Carolina & 43757 & 33 \\
Lancaster & South Carolina & Union & North Carolina & 14117 & 38 \\
Spartanburg & South Carolina & Polk & North Carolina & 12227 & 41 \\
Cherokee & South Carolina & Cleveland & North Carolina & 6703 & 25 \\
Chesterfield & South Carolina & Anson & North Carolina & 6595 & 32 \\
Marlboro & South Carolina & Scotland & North Carolina & 5109 & 21 \\
Dillon & South Carolina & Robeson & North Carolina & 4497 & 37 \\
\hline
\end{tabular}
\end{scriptsize}
\end{center}
\caption{Democrats (150,000 movements)}
\label{tab:D_150000}\end{table}
% \vspace{0.5cm}

\begin{table}[H]
\begin{center}
\begin{scriptsize}
\begin{tabular}{|c|c|c|c|c|c|}
\hline
\multicolumn{2}{|c|}{\textbf{Origin}} & \multicolumn{2}{c|}{\textbf{Destination}} & {\centering \textbf{Movement}} & {\centering \textbf{Distance}} \\ \cline{1-4}
\textbf{County} & \textbf{State} & \textbf{County} & \textbf{State} &  &  \\ \hline
& & & \textbf{Arizona} & \textbf{18611} & \\
\cdashline{1-6}
Imperial & California & Yuma & Arizona & 12600 & 96 \\
Mckinley & New Mexico & Apache & Arizona & 6010 & 63 \\
\hline
& & & \textbf{Georgia} & \textbf{25986} & \\
\cdashline{1-6}
Russell & Alabama & Muscogee & Georgia & 8179 & 16 \\
Chambers & Alabama & Troup & Georgia & 7452 & 27 \\
Randolph & Alabama & Heard & Georgia & 6569 & 29 \\
Nassau & Florida & Camden & Georgia & 3312 & 19 \\
Cleburne & Alabama & Haralson & Georgia & 472 & 33 \\
\hline
& & & \textbf{Nevada} & \textbf{40688} & \\
\cdashline{1-6}
Nevada & California & Carson & Nevada & 19070 & 90 \\
El Dorado & California & Douglas & Nevada & 16575 & 91 \\
Inyo & California & Esmeralda & Nevada & 4056 & 78 \\
Sierra & California & Washoe & Nevada & 854 & 47 \\
Alpine & California & Douglas & Nevada & 130 & 26 \\
\hline
& & & \textbf{New Hampshire} & \textbf{54773} & \\
\cdashline{1-6}
Essex & Massachusetts & Hillsborough & New Hampshire & 54773 & 45 \\
\hline
& & & \textbf{North Carolina} & \textbf{50753} & \\
\cdashline{1-6}
Greene & Tennessee & Madison & North Carolina & 16634 & 43 \\
Carter & Tennessee & Avery & North Carolina & 15183 & 37 \\
Washington & Tennessee & Yancey & North Carolina & 9479 & 48 \\
Unicoi & Tennessee & Yancey & North Carolina & 4883 & 29 \\
Johnson & Tennessee & Watauga & North Carolina & 4573 & 30 \\
\hline
& & & \textbf{Wisconsin} & \textbf{9186} & \\
\cdashline{1-6}
Lake & Illinois & Kenosha & Wisconsin & 9186 & 27 \\
\hline
\end{tabular}
\end{scriptsize}
\end{center}
\caption{Republicans (200,000 movements)}
\label{tab:R_200000}\end{table}
% \vspace{0.5cm}

\begin{table}[H]
\begin{center}
\begin{scriptsize}
\begin{tabular}{|c|c|c|c|c|c|}
\hline
\multicolumn{2}{|c|}{\textbf{Origin}} & \multicolumn{2}{c|}{\textbf{Destination}} & {\centering \textbf{Movement}} & {\centering \textbf{Distance}} \\ \cline{1-4}
\textbf{County} & \textbf{State} & \textbf{County} & \textbf{State} &  &  \\ \hline
& & & \textbf{Georgia} & \textbf{19548} & \\
\cdashline{1-6}
Aiken & South Carolina & Richmond & Georgia & 11745 & 28 \\
Jasper & South Carolina & Effingham & Georgia & 5293 & 26 \\
Mccormick & South Carolina & Lincoln & Georgia & 2510 & 19 \\
\hline
& & & \textbf{Nevada} & \textbf{48509} & \\
\cdashline{1-6}
Nevada & California & Carson & Nevada & 25514 & 90 \\
El Dorado & California & Douglas & Nevada & 18751 & 91 \\
Inyo & California & Esmeralda & Nevada & 3344 & 78 \\
Sierra & California & Washoe & Nevada & 555 & 47 \\
Alpine & California & Douglas & Nevada & 343 & 26 \\
\hline
& & & \textbf{New Hampshire} & \textbf{6441} & \\
\cdashline{1-6}
Essex & Massachusetts & Hillsborough & New Hampshire & 6441 & 45 \\
\hline
& & & \textbf{North Carolina} & \textbf{125501} & \\
\cdashline{1-6}
York & South Carolina & Mecklenburg & North Carolina & 43757 & 33 \\
Spartanburg & South Carolina & Polk & North Carolina & 43326 & 41 \\
Lancaster & South Carolina & Union & North Carolina & 14117 & 38 \\
Cherokee & South Carolina & Cleveland & North Carolina & 6703 & 25 \\
Chesterfield & South Carolina & Anson & North Carolina & 6595 & 32 \\
Marlboro & South Carolina & Scotland & North Carolina & 5109 & 21 \\
Dillon & South Carolina & Robeson & North Carolina & 4497 & 37 \\
Greenville & South Carolina & Polk & North Carolina & 1393 & 45 \\
\hline
\end{tabular}
\end{scriptsize}
\end{center}
\caption{Democrats (200,000 movements)}
\label{tab:D_200000}\end{table}
% \vspace{0.5cm}

\begin{table}[H]
\begin{center}
\begin{scriptsize}
\begin{tabular}{|c|c|c|c|c|c|}
\hline
\multicolumn{2}{|c|}{\textbf{Origin}} & \multicolumn{2}{c|}{\textbf{Destination}} & {\centering \textbf{Movement}} & {\centering \textbf{Distance}} \\ \cline{1-4}
\textbf{County} & \textbf{State} & \textbf{County} & \textbf{State} &  &  \\ \hline
& & & \textbf{Arizona} & \textbf{18611} & \\
\cdashline{1-6}
Imperial & California & Yuma & Arizona & 13965 & 96 \\
Mckinley & New Mexico & Apache & Arizona & 4645 & 63 \\
\hline
& & & \textbf{Georgia} & \textbf{24987} & \\
\cdashline{1-6}
Hamilton & Tennessee & Catoosa & Georgia & 24987 & 15 \\
\hline
& & & \textbf{Michigan} & \textbf{2773} & \\
\cdashline{1-6}
Lagrange & Indiana & St. Joseph & Michigan & 2773 & 25 \\
\hline
& & & \textbf{Nevada} & \textbf{46591} & \\
\cdashline{1-6}
El Dorado & California & Douglas & Nevada & 22478 & 91 \\
Nevada & California & Carson & Nevada & 19070 & 90 \\
Inyo & California & Esmeralda & Nevada & 4056 & 78 \\
Sierra & California & Washoe & Nevada & 854 & 47 \\
Alpine & California & Douglas & Nevada & 130 & 26 \\
\hline
& & & \textbf{New Hampshire} & \textbf{54773} & \\
\cdashline{1-6}
Essex & Massachusetts & Hillsborough & New Hampshire & 54773 & 45 \\
\hline
& & & \textbf{North Carolina} & \textbf{62719} & \\
\cdashline{1-6}
York & South Carolina & Mecklenburg & North Carolina & 17867 & 33 \\
Cherokee & South Carolina & Cleveland & North Carolina & 12395 & 25 \\
Chesterfield & South Carolina & Anson & North Carolina & 7679 & 32 \\
Carter & Tennessee & Avery & North Carolina & 5458 & 37 \\
Grayson & Virginia & Alleghany & North Carolina & 4907 & 16 \\
Unicoi & Tennessee & Yancey & North Carolina & 4883 & 29 \\
Johnson & Tennessee & Watauga & North Carolina & 4573 & 30 \\
Marlboro & South Carolina & Scotland & North Carolina & 3507 & 21 \\
Danville & Virginia & Caswell & North Carolina & 1446 & 19 \\
\hline
& & & \textbf{Pennsylvania} & \textbf{2215} & \\
\cdashline{1-6}
New Castle & Delaware & Chester & Pennsylvania & 2215 & 36 \\
\hline
& & & \textbf{Wisconsin} & \textbf{37327} & \\
\cdashline{1-6}
Lake & Illinois & Kenosha & Wisconsin & 37327 & 27 \\
\hline
\end{tabular}
\end{scriptsize}
\end{center}
\caption{Republicans (250,000 movements)}
\label{tab:R_250000}\end{table}
% \vspace{0.5cm}

\begin{table}[H]
\begin{center}
\begin{scriptsize}
\begin{tabular}{|c|c|c|c|c|c|}
\hline
\multicolumn{2}{|c|}{\textbf{Origin}} & \multicolumn{2}{c|}{\textbf{Destination}} & {\centering \textbf{Movement}} & {\centering \textbf{Distance}} \\ \cline{1-4}
\textbf{County} & \textbf{State} & \textbf{County} & \textbf{State} &  &  \\ \hline
& & & \textbf{Georgia} & \textbf{19229} & \\
\cdashline{1-6}
Nassau & Florida & Camden & Georgia & 11044 & 19 \\
Gadsden & Florida & Decatur & Georgia & 8184 & 34 \\
\hline
& & & \textbf{Nevada} & \textbf{59728} & \\
\cdashline{1-6}
El Dorado & California & Douglas & Nevada & 29970 & 91 \\
Nevada & California & Carson & Nevada & 25514 & 90 \\
Inyo & California & Esmeralda & Nevada & 3344 & 78 \\
Sierra & California & Washoe & Nevada & 555 & 47 \\
Alpine & California & Douglas & Nevada & 343 & 26 \\
\hline
& & & \textbf{New Hampshire} & \textbf{8883} & \\
\cdashline{1-6}
Essex & Massachusetts & Hillsborough & New Hampshire & 8883 & 45 \\
\hline
& & & \textbf{North Carolina} & \textbf{162158} & \\
\cdashline{1-6}
York & South Carolina & Mecklenburg & North Carolina & 43757 & 33 \\
Spartanburg & South Carolina & Polk & North Carolina & 43326 & 41 \\
Greenville & South Carolina & Polk & North Carolina & 38051 & 45 \\
Lancaster & South Carolina & Union & North Carolina & 14117 & 38 \\
Cherokee & South Carolina & Cleveland & North Carolina & 6703 & 25 \\
Chesterfield & South Carolina & Anson & North Carolina & 6595 & 32 \\
Marlboro & South Carolina & Scotland & North Carolina & 5109 & 21 \\
Dillon & South Carolina & Robeson & North Carolina & 4497 & 37 \\
\hline
\end{tabular}
\end{scriptsize}
\end{center}
\caption{Democrats (250,000 movements)}
\label{tab:D_250000}\end{table}
%\vspace{0.5cm}

% \end{APPENDIX}

 % \end{APPENDICES}

% \bibliographystyle{informs2014} % outcomment this and next line in Case 1
% \bibliography{references} % if more than one, comma separated

\end{document}